# DESIGN OF THE LBNE BEAMLINE*

V. Papadimitriou[#], R. Andrews, J. Hylen, T. Kobilarcik, A. Marchionni, C. D. Moore, P. Schlabach, S. Tariq, Fermilab, Batavia, IL 60510, USA

*Abstract*

The Long Baseline Neutrino Experiment (LBNE) will utilize a beamline facility located at Fermilab to carry out a compelling research program in neutrino physics. The facility will aim a wide band beam of neutrinos toward a detector placed at the Sanford Underground Research Facility in South Dakota, about 1,300 km away. The main elements of the facility are a primary proton beamline and a neutrino beamline. The primary proton beam (60-120 GeV) will be extracted from the MI-10 section of Fermilab's Main Injector. Neutrinos are produced after the protons hit a solid target and produce mesons which are sign selected and subsequently focused by a set of magnetic horns into a 204 m long decay pipe where they decay mostly into muons and neutrinos. The parameters of the facility were determined taking into account the physics goals, spacial and radiological constraints and the experience gained by operating the NuMI facility at Fermilab. The initial beam power is expected to be ~1.2 MW, however the facility is designed to be upgradeable for 2.3 MW operation. We discuss here the status of the design and the associated challenges.

## INTRODUCTION

The Beamline is a central component of LBNE and its driving physics considerations are the long baseline neutrino oscillation analyses. On January 8, 2010 the Department of Energy approved the "Mission Need" (CD-0) for LBNE and on December 10, 2012 the conceptual design (CD-1).

The beamline facility is expected to be fully contained within Fermilab property. The primary proton beam, in the energy range of 60-120 GeV, will be extracted from the Main Injector's (MI) [1] MI-10 section using "single-turn" extraction. For 120 GeV operation and with the MI upgrades implemented for the NOvA experiment [2] as well as with the expected implementation of the accelerator Proton Improvement Plan, phase II (PIP-II) [3], the fast, single turn extraction will deliver all the protons ($7.5 \times 10^{13}$) in one MI machine cycle (1.2 sec) to the LBNE target in 10 μs. The beam power is expected to be 1.2 MW in the energy range of 80 to 120 GeV [3]. The charged mesons produced by the interaction of the protons with the target are sign selected and focused by two magnetic horns into the decay pipe towards the far detector. These mesons are short-lived and decay into muons and neutrinos. At the end of the decay region, an absorber pile is needed to remove the residual hadrons remaining at the end of the decay pipe. The neutrino beam is aimed 4850 ft underground at the Sanford Underground Research Facility in South Dakota (SURF), about 1300 km away.

A wide band neutrino beam is needed to cover the first and second neutrino oscillation maxima, which for a 1300 km baseline are expected to be approximately at 2.4 and 0.8 GeV. The beam must provide a high neutrino flux at the energies bounded by the oscillation peaks and we are therefore optimizing the beamline design for neutrino energies in the area 0.5 to 5 GeV. The initial operation of the facility will be at a beam power incident on the production target of 1.2 MW, however some of the initial implementation will have to be done in such a manner that operation at 2.3 MW can be achieved without retrofitting. Such a higher beam power is expected to become available in the future with additional improvements in the Fermilab accelerator complex [4]. In general, components of the LBNE beamline system which cannot be replaced or easily modified after substantial irradiation at 1.2 MW operation are being designed for 2.3 MW. Examples of such components are the shielding of the target chase and the decay pipe, and the hadron absorber.

The LBNE Beamline design has to implement as well stringent limits on the radiological protection of the environment, workers and members of the public. The relevant radiological concerns, prompt dose, residual dose, air activation and water activation have been extensively modeled and the results are incorporated in the system design. A most important aspect of modeling at the present design stage is the determination of the necessary shielding thickness and composition in order to protect the ground water and the public and to control air emissions.

This paper is a snapshot of the present status of the design as exemplified in the Conceptual Design Report [5] and the recently published LBNE science document [6].

## STATUS OF THE DESIGN

Figure 1 shows a longitudinal section of the LBNE beamline facility. At MI-10 there is no existing extraction enclosure and we are minimizing the impact on the MI by introducing a 15.6 m long beam carrier pipe to transport the beam through the MI tunnel wall into the new LBNE enclosure. The extraction and transport components send the proton beam through a man-made embankment/hill whose apex is ~18 m and the footprint is ~21,370 m². The beam then will be bent downward toward a target located at grade level. The overall bend of the proton beam is 7.2° westward and 5.8° downward to establish the final trajectory toward the far detector.

___________________________________________
*Work supported by DOE, contract No. DE-AC02-07CH11359
#vaia@fnal.gov

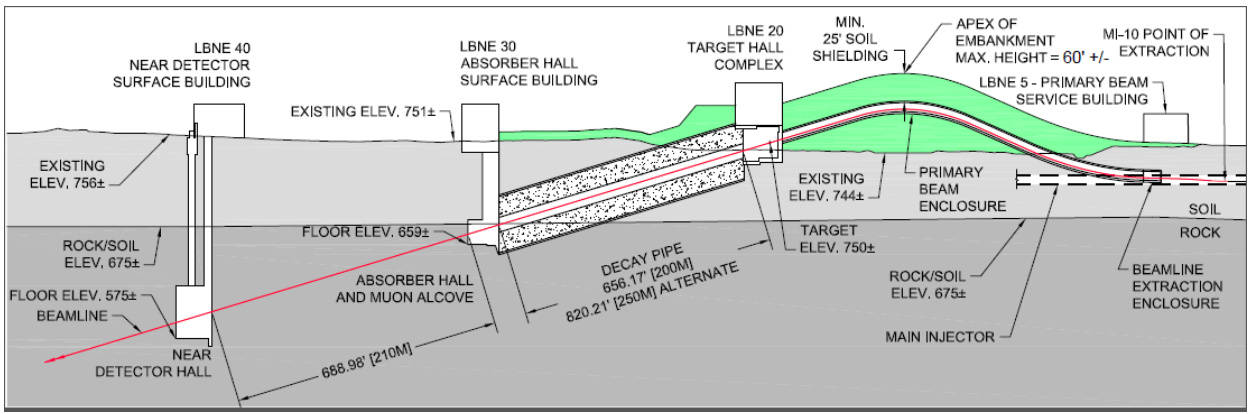

Figure 1: Longitudinal section of the LBNE beamline facility at Fermilab. The beam comes from the right, the protons being extracted from the MI-10 straight section of the MI.

In this shallow beamline design, because of the presence of a local aquifer at and near the top of the rock surface, an engineered geomembrane barrier and drainage system between the shielding and the environment prevents the contamination of groundwater from radionuclides. The decay pipe shielding thickness has been determined to be 5.6 m of concrete (see Fig. 2).

## Beamline Scope

The LBNE Beamline scope includes a primary (proton) beamline, a neutrino beamline and associated conventional facilities. The Primary Beamline elements necessary for extraction and transport include vacuum pipes, dipole, quadrupole, corrector, kicker, lambertson and C magnets and beam monitoring equipment (Beam-Position Monitors, Beam-Loss Monitors, Beam-Profile Monitors and Beam-Intensity monitors). The magnets are conventional, MI design, and the magnet power supplies are a mixture of new, MI design power supplies or refurbished Tevatron power supplies. The beam optics accommodates a range of spot sizes on the target in the energy range of interest and for beam power up to 2.3 MW, and the beam transport is expected to take place with negligible losses.

The neutrino beamline includes in order of placement (1) a beryllium window that seals off and separates the evacuated primary beamline from the neutrino beamline, (2) a baffle collimator assembly to protect the target and the horns from beam accidents, (3) a target, (4) two magnetic horns. These elements are all located inside a heavily shielded, air-filled, air/water-cooled vault, called the target chase (see Fig. 2), that is isolated from the decay pipe at its downstream end by a replaceable, thin, metallic window. A 204 m long, 4m in diameter helium-filled, air-cooled decay pipe follows which allows pions and kaons to decay to neutrinos. At the end of it is the hadron absorber which is a specially designed pile of aluminum, steel and concrete blocks, some of them water cooled, which must contain the energy of the particles that exit the decay pipe. Radiation damage, cooling of elements, radionuclide mitigation, remote handling and storage of radioactive components are essential considerations for the conceptual design of the neutrino beamline.

## Target and horns

The reference target design for LBNE is an upgraded version of the NuMI low-energy target that was used for eight years to deliver beam to the NuMI experiment. The target core is made of POCO ZXF-5Q graphite segmented into short rectangular segments oriented vertically, with the short direction transverse to the beam. The 95 cm long target consists of 47 20 mm long segments and it is water cooled. Until January 2014 (when we were still expecting to start operating LBNE at 700 kW) the segments were 7.4 mm in width. Also, in order to capture low energy pions and kaons from the target at large production angles, 60 cm of the target assembly was inserted into the first horn's magnetic lens. The horns are of NuMI design with their inner conductors having a double parabolic shape. Although they were designed for 200 kA we re-evaluated them and determined they can run at 230 kA. The plan was to power the LBNE horns by using the NuMI power supply with current pulse width of 2.1 ms.

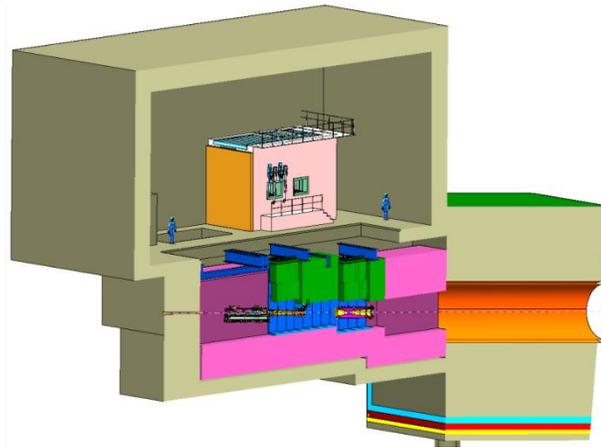

Figure 2: Schematic of the upstream portion of the LBNE neutrino beamline showing the major components of the neutrino beam. The target chase bulk steel shielding is shown in magenda. Inside the chase from left to right one can see the horn-protection baffle and the target mounted on a carrier and the two focusing horns. The decay pipe follows (in orange). The beige areas indicate concrete shielding.

*The 1.2 MW challenge*

One of the challenges for the LBNE Beamline is the recent requirement to have the Beamline ready to accept 1.2 MW of beam power on day one of LBNE operations. Our approach has been to check if modest modifications to the CD-1 designs can achieve this. We attempted to reduce stress on the target by increasing the beam sigma from 1.3 to 1.7 mm. This led to an increase of the graphite segment width to 10 mm (see Fig. 4) which subsequently led to retraction of the target by 10 cm. The thermal and stress analysis for 120 GeV operations indicates excellent safety factors (>7.5) for the graphite segments, the beryllium container and the beryllium window of the target assembly. The safety factor for the titanium water lines is ~2.5.

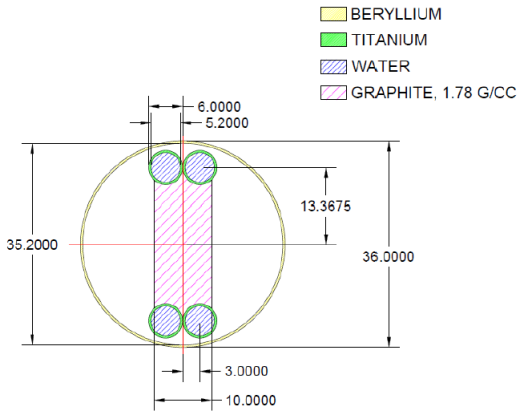

Figure 4: Cross section of the LBNE 1.2 MW target. All dimensions are given in millimeters.

For the horns, we attempted to reduce the joule heating to allow space for more beam heating by using a new power supply with shorter current pulse width (0.8 ms). This change allowed for the current to remain at 230 kA. Thermal and stress analysis for horn 1 pointed to the need to round off the inner conductor's transition from parabola to neck and to move further upstream the upstream weld. The horn safety factors for 120 GeV operation are now greater than 3.5 for all areas of horn 1. Our next step is to perform the same analyses for 80 GeV operation for both the target and the horns.

Figure 5 shows the distribution of unoscillated $\nu_\mu$ charge current events per GeV, for different proton energies, for three years of operation at 1.2 MW and with a 34 kton liquid argon TPC far detector. We notice that at lower proton energies we get more events in the neutrino energy range corresponding to the 2$^{nd}$ oscillation maximum.

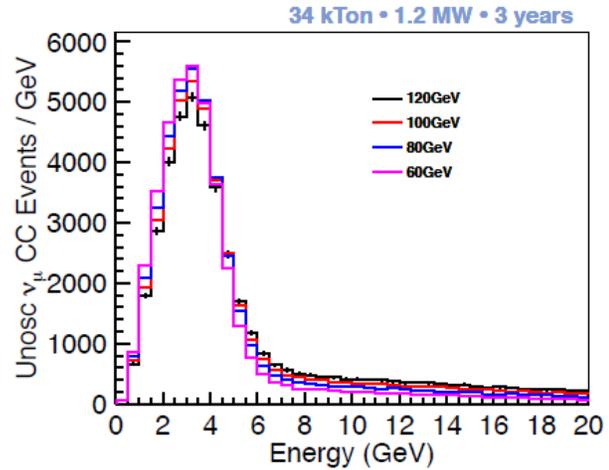

Figure 5: Unoscillated $\nu_\mu$ charge current events per GeV as a function of neutrino energy for different proton energies.

## CONCLUSION

The LBNE Beamline team developed a conceptual design for the Beamline and went through a successful DOE review that resulted in CD-1 approval in December 2012. We are now in the process of advancing this design towards baselining while addressing at the same time a new challenge of having the beamline ready to accept beam power of 1.2 MW at day one of LBNE operations. We have so far successfully responded to this challenge with small modifications to the CD-1 target and horn designs.

## ACKNOWLEDGMENT

We would like to thank all the members of the LBNE Beamline team for their numerous contributions towards developing and costing the Beamline conceptual design, and advancing it in all areas towards baselining.